\documentclass[french]{pfia} 

\usepackage[T1]{fontenc}
\usepackage[utf8]{inputenc}
\usepackage{comment}
\usepackage{subcaption}
\usepackage{enumitem}
\usepackage{graphicx}
\usepackage[dvipsnames]{xcolor}
 
\usepackage{url}
\definecolor{afiablue}{RGB}{61,159,207}
\definecolor{afiared}{RGB}{167,75,68}
\definecolor{afialightblue}{RGB}{158,193,232}

\newcommand{\field}[1]{\textcolor{afiablue}{#1}}
\newcommand{\val}[1]{\textsf{\textcolor{afialightblue}{#1}}}

\def\etc.{etc.\spacefactor=\the\sfcode`\c}

\newcommand{\fig}[1]{Figure~\ref{#1}}

\newcommand{\couleur}[2]{{\color{#1}#2}}

\newcommand{\sic}[1]{\og{}{\it #1 }\fg{}}

\newcommand{\isoKM}{{\sc ISO~30401:2018 }}
\newcommand{\isoQ}{{\sc ISO~9001:2015 }}

\newcommand{\guilli}[1]{\og{}{\it #1}\fg{}}
\newcommand{\lepdca}{{\couleur{PineGreen}{{\bf PDCA }}}}
\newcommand{\leseci}{{\couleur{RoyalBlue}{{\bf SECI }}}}
\newcommand{\pdca}[2]{{\couleur{PineGreen}{{\bf #1}{\it #2}}}}
\newcommand{\seci}[2]{{\couleur{RoyalBlue}{{\bf #1}{\it #2}}}}

\newcommand{\ittguilli}[2]{\item[$\triangleright$]\guilli{{\couleur{RoyalBlue}{#1}}}{ : #2}}
\newcommand{\ittguillj}[2]{\item[$\triangleright$]{\couleur{RoyalBlue}{#1} }{#2}}

\newcommand{\ProcPilotage}[1]{{\couleur{Red} {#1}}}
\newcommand{\ProcRealisation}[1]{{\couleur{Blue} {#1}}}
\newcommand{\ProcSupport}[1]{{\couleur{Green} {#1}}}

\newcommand{\guil}[1]{%
 \ifthenelse{\isempty{#1}}%
    {}
    {\og{}#1\fg{}}
                    }

\title{\textbf{Integrating an ISO~30401-compliant Knowledge Management System with the processes of an Integrated Management System}}

\author{Patrick Prieur \fup{1} \& Aline Belloni\fup{1}\\[6pt]
\fup{1} Ardans SAS,\\ 6 rue Jean Pierre Timbaud, \sic{Le Campus} Bâtiment B1, 78180 Montigny-le-Bretonneux, France\\
 \{pprieur, abelloni\}@ardans.fr $\bullet$        \url{https://www.ardans.fr}\\
 }

\date{July 3, 2025}

\begin{document}

\maketitle

\begin{abstract}
With the evolution of process approaches within organizations, the increasing importance of quality management systems (like ISO 9001) and the recent introduction of ISO 30401 for knowledge management, we examine how these different elements converge within the framework of an Integrated Management System. The article specifically demonstrates how a knowledge management system can be implemented by deploying the mechanisms of the \leseci ~model through the steps of the \lepdca ~cycle as applied in the processes of the integrated management system.
\end{abstract}



\section{Introduction}
\isoQ standard \cite{iso9000} defines a process as a  \sic{set of interrelated or interacting activities that use inputs to deliver an intended result}. The process approach within an organization then involves identifying, describing and grouping activities and tasks that contribute to achieving the organization's results into processes. These processes are then combined into an Integrated Management System, interlinking operational (realization), steering (management) and support levels according to the systemic vision developed by Le Moigne \cite{LeMoigne78} or de Rosnay \cite{deRosnay75} and adopted by ISO 9000:2008 \cite{iso2008}. Whether certified or simply "aligned" with this standard, many organizations use the process approach as an essential framework to ensure "quality", i.e. aligning tasks with their strategic objectives (effectiveness) and ensuring the efficiency of their employees' work and workflows.

Introduced by \isoQ as a new requirement for quality management systems, the knowledge management "sub-system" within an organization was further detailed in 2018 with the publication of \isoKM \cite{iso30401}. This standard belongs to the family of Management System Standards (MSS) developed to offer a harmonized normative framework for the most common processes found in organizations\footnote{ The complete list of Management System Standards, both sectoral and non-sectoral, is available on the ISO website \cite{isoMgt}. Additionally, ISO has published a manual to help organizations integrate the requirements of multiple MSS into their own management systems \cite{iumss2018}.}. In its current version (with a revision currently underway), \isoKM prescribes four activities dedicated to knowledge development (see chapter 4.4.2 of the standard): acquisition, application, retention and the management of obsolete or non-conforming knowledge. It also outlines four activities and behaviours related to knowledge conveyance and transformation (see chapter 4.4.3 of the standard): human interaction, representation, combination, Internalization and learning. We can observe that these latter points are directly inspired by Nonaka and Takeuchi's \leseci model \cite{Nonaka95} and \cite{Nonaka19}, which we will discuss further below.

As expert practitioners in implementing \isoKM (see \cite{ArdansIC2023}), we are regularly faced with the need to explain to the teams we support how the knowledge development, transfer and transformation activities described in \isoKM integrate with their existing operational processes which serve as their reference point. Through these exercises, we have become convinced that an organization genuinely practising the continuous improvement cycle advocated by \isoQ is in a favourable position to easily implement the mechanisms of the \leseci model.
Based on our experience implementing \isoKM over the past six years in French public industrial and commercial establishments (EPIC) and large private sector companies, this article will first quickly explore the process modelling principles introduced by \isoQ and the \leseci model. Then, it will present our current thinking on how to articulate a knowledge management system compliant with \isoKM with the other processes of an integrated management system.

\section{Conceptual Framework for Process Modelling}
The process approach provides a process map that offers a global view of the flows between processes and a detailed representation of the activities and tasks within each process.

\begin{figure}[ht]
    \centering
    \includegraphics[width=\textwidth/2]{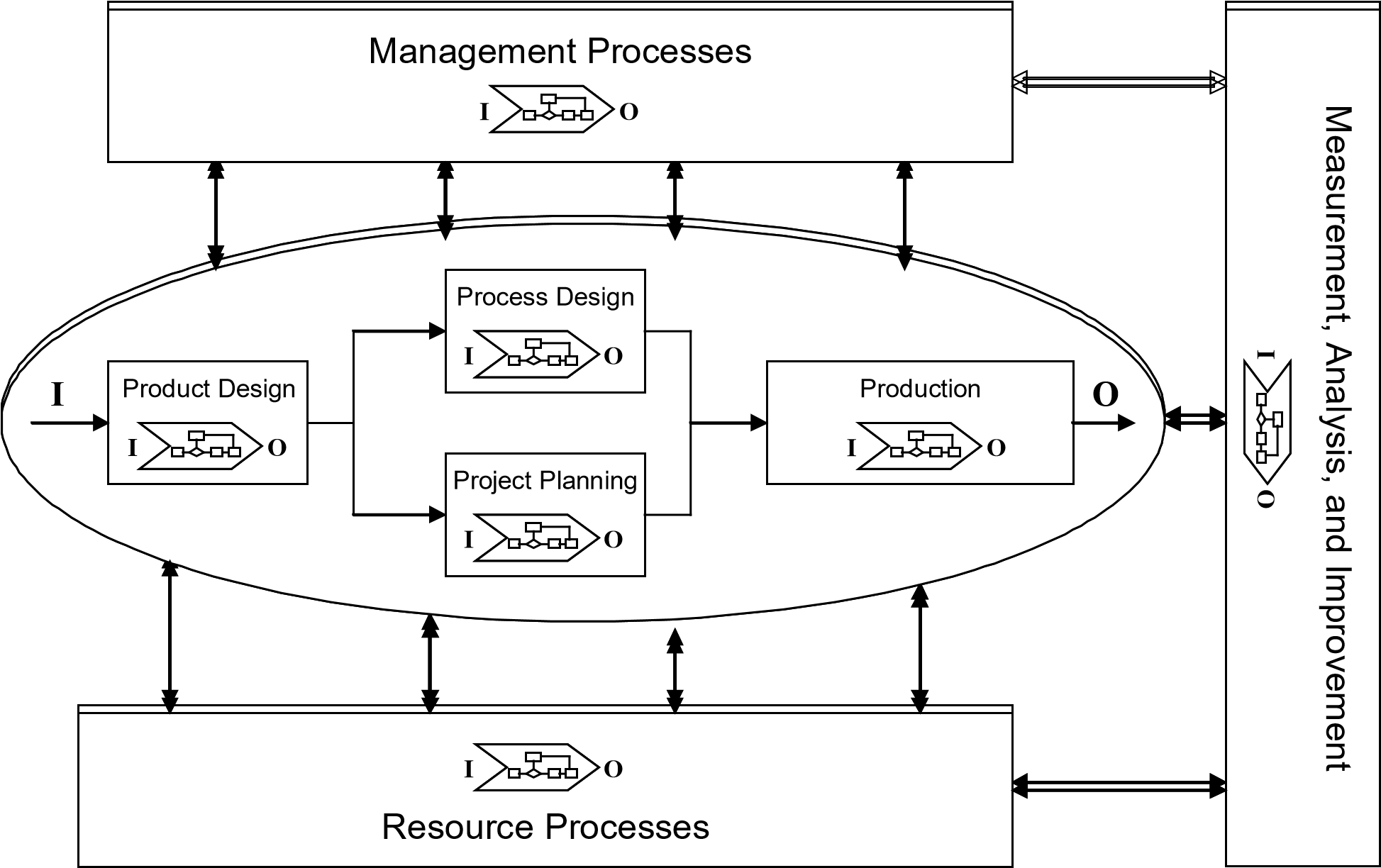}
    \caption{"Example of a process sequence ans its interactions" from \cite{iso2008}}
    \label{Fig_ArtApia_01}
\end{figure}

\begin{figure}[ht]
    \centering
    \includegraphics[width=\textwidth/2]{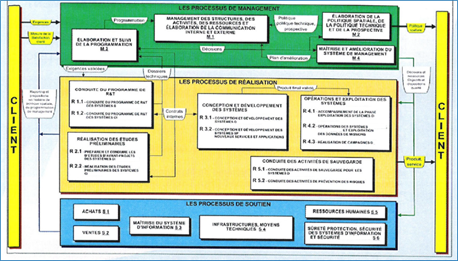}
    \caption{An ISO 9001-compliant process map of a French public industrial and commercial establishment}
    \label{Fig_ArtApia_02}
\end{figure}

The process approach application guide \cite{iso2008}, published alongside the ISO 9000:2008 family of standards, suggested creating this mapping based on a categorization into process families, depicted in \fig{Fig_ArtApia_01} :
\begin{itemize}
    \item Management Processes (or steering or leadership processes)
    \item Realization Processes (or operational processes, core business processes). These processes form the organization's value chain
    \item Support Processes (or resource processes)
    \item Measurement, Analysis and Improvement Processes
\end{itemize}
Measurement processes are often documented as activities directly within the management, realization and support processes they concern. In contrast, analysis and improvement processes are frequently treated as independent processes that interact with the other processes, taking the results of measurement activities as input and generating improvement instructions for these other processes as output. \fig{Fig_ArtApia_02} corresponds to an example of the implementation of this systemic approach observed in a French public industrial and commercial establishment (EPIC).

\begin{figure}[ht]
    \centering
    \includegraphics[width=\textwidth/2]{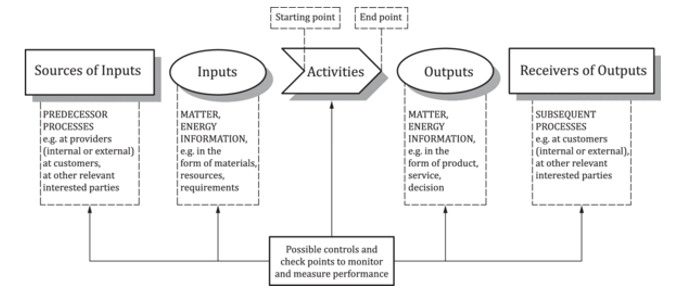}
    \caption{\isoQ schematic representation of the elements of a single process, from \cite{iso9001} : }
    \label{Fig_ArtApia_04}
\end{figure}

At the heart of \isoQ  process mapping is the "Activity" model (\fig{Fig_ArtApia_04}), largely inspired by well-known system modelling methodologies such as SADT \cite{galinier1989} or \cite{IDEF0}. One or more activities describe a process and each of these activities are, in turn, described by more elementary tasks.

In relation to this diagram, a further step can be taken by using the famous "Turtle Diagram," whose components were introduced by Crosby as early as 1979 \cite{Crosby1979}. This diagram specifies the elements that should be detailed to fully describe a process or an activity:
\begin{itemize}
\item The Shell: The process itself (to be detailed by its constituent activities)
\item The Head: The inputs to the process, that are going to be transformed.
\item The Tail: The outputs of the process (products or services)
\item The Legs: The required resources (who, with what), the methods (how) and the performance indicators (how much).
\end{itemize}
\begin{figure}[ht]
    \centering
    \includegraphics[width=\textwidth/2]{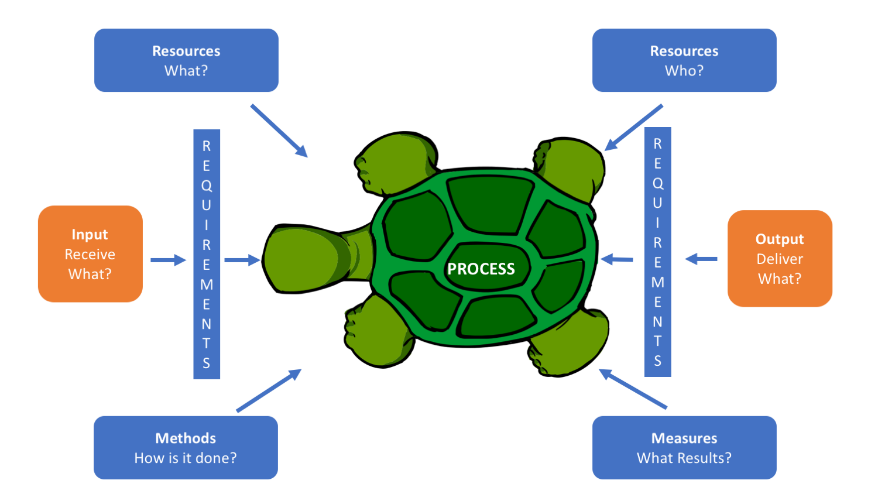}
    \caption{\sic{How to use Turtle Diagrams} according to IATF16949 Standards Store}
    \label{ArtApia_Turtle}
\end{figure}

\begin{figure}[ht]
    \centering
    \includegraphics[width=\textwidth/2]{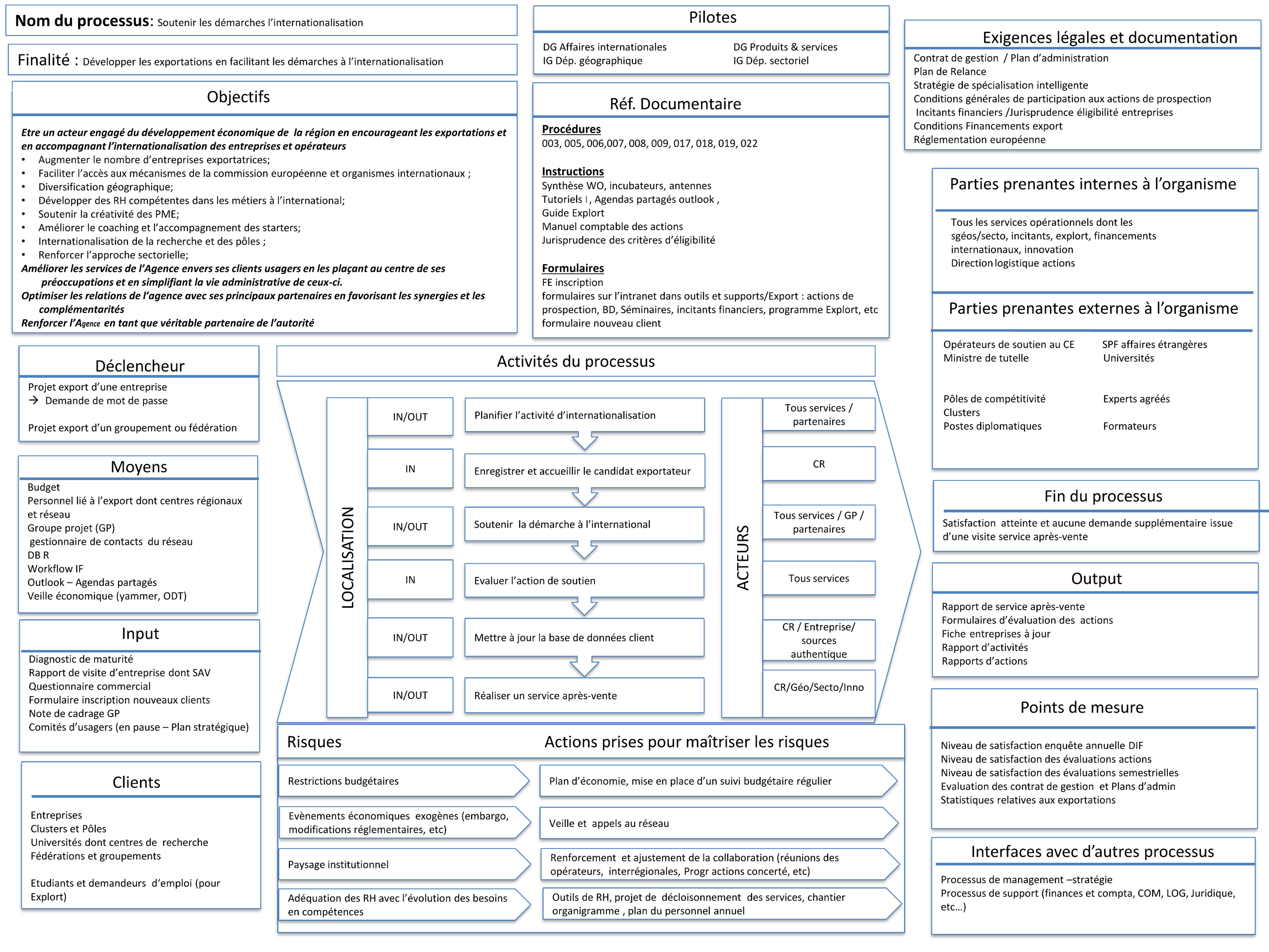}
    \caption{Turtle diagram application example in a state agency}
    \label{ArtApia_05}
\end{figure}

\fig{ArtApia_Turtle} illustrates the Turtle Diagram\footnote{https://16949store.com/articles/how-to-use-turtle-diagrams}  and \fig{ArtApia_05} gives an application example over a process in a state agency. 
The value of this type of diagram, from a knowledge management perspective, is that it {\it de facto}  maps the explicit knowledge required for process execution, without necessarily detailing it very deeply. This generally yields an overview of knowledge related to inputs, outputs and environmental constraints (the "what?"), as well as the know-how needed to transform inputs into outputs (the "who?", "with what?", "how?").

\section{ISO 9001:2015: Beyond simply "Doing" a process : The PDCA Cycle associated with process steering and supporting}\label{sect:DoPDCA}

The \lepdca cycle, also known as the "Deming Wheel" \cite{Deming}, is the continuous improvement cycle adopted by ISO 9000 since its 2000 version. Its "PLAN-DO-CHECK-ACT" principle involves planning an activity ("Plan"), then doing it ("Do"), which means implementing the planned activity. Next, you check ("Check") the results obtained and act ("Act") by taking appropriate measures to adjust the subsequent cycle based on the observations.

Building on the 2008 version, \isoQ makes it a requirement for organizations to adopt a holistic approach for their process modelling. It emphasizes that describing a process or an activity is not just about detailing the "processing" or "Doing" aspect when realizing something. Instead, it's about describing the entire system that contributes to that realization. This means adding the description of the continuous improvement loop for the realization process, as well as the steering (\sic{leadership}) and \sic{support} processes, as previously discussed within the systemic vision of the process approach:
\begin{itemize}
\item Realization : Plan – Do – Check – Act (or Adjust)
\item Lead (or steer or \sic{Manage})
\item Support
\end{itemize}

Based on this principle, \isoQ proposes the following generic representation of any management system (\fig{Fig_ArtApia_03}):

\begin{figure}[ht]
    \centering
    \includegraphics[width=\textwidth/2]{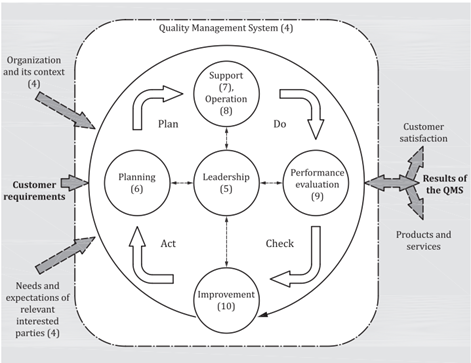}
    \caption{\sic{Representation of the structure of this International Standard in the \lepdca cycle} from \cite{iso9001} }
    \label{Fig_ArtApia_03}
\end{figure}
Following the fractal principle of systemic modelling, the continuous improvement cycle is found at all levels of a process, from the broadest overview to the most granular step (task). We should note, however, that the degree of formalization in applying \lepdca varies accordingly: from generally very formalized and detailed at the overall process level\dots{} to often completely informal at the most elementary level of a task executed by an individual.
\begin{figure}[ht]
    \centering
    \includegraphics[width=\textwidth/2]{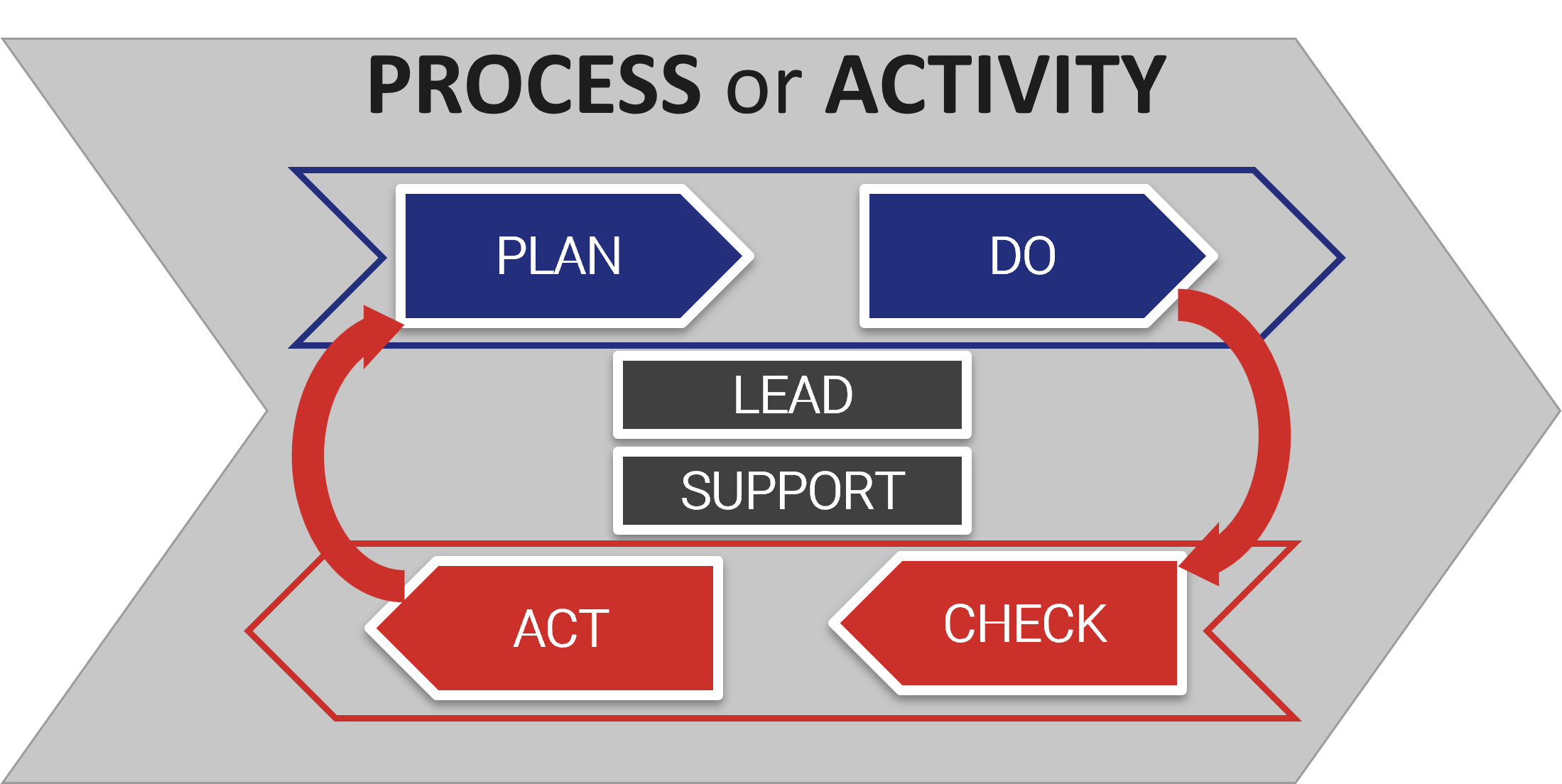}
    \caption{Reformulation de la représentation générique}
    \label{ArtApiaProcAct}
\end{figure}
Adopting the generic representation of a process, sub-process, or activity given in \fig{ArtApiaProcAct}, for the sake of global process mapping we can continue grouping all or part of the Leadership and Support activities into processes that belong to the "Management" and "support" families of the systemic approach mentioned earlier: 

--

\begin{tabular}{p{0.5\textwidth}}
$\triangleright$ \ProcRealisation{The "Leadership" activity of a realization process}\\ becomes \ProcPilotage{the "Realization" activity of a Management Process}.\\
$\triangleright$ \ProcRealisation{The "Support" activity of a realization process}\\ becomes \ProcSupport{the "Realization" activity of a Support Process}.\\
\end{tabular}
--

\begin{figure}[ht]
    \centering
    \includegraphics[width=\textwidth/2]{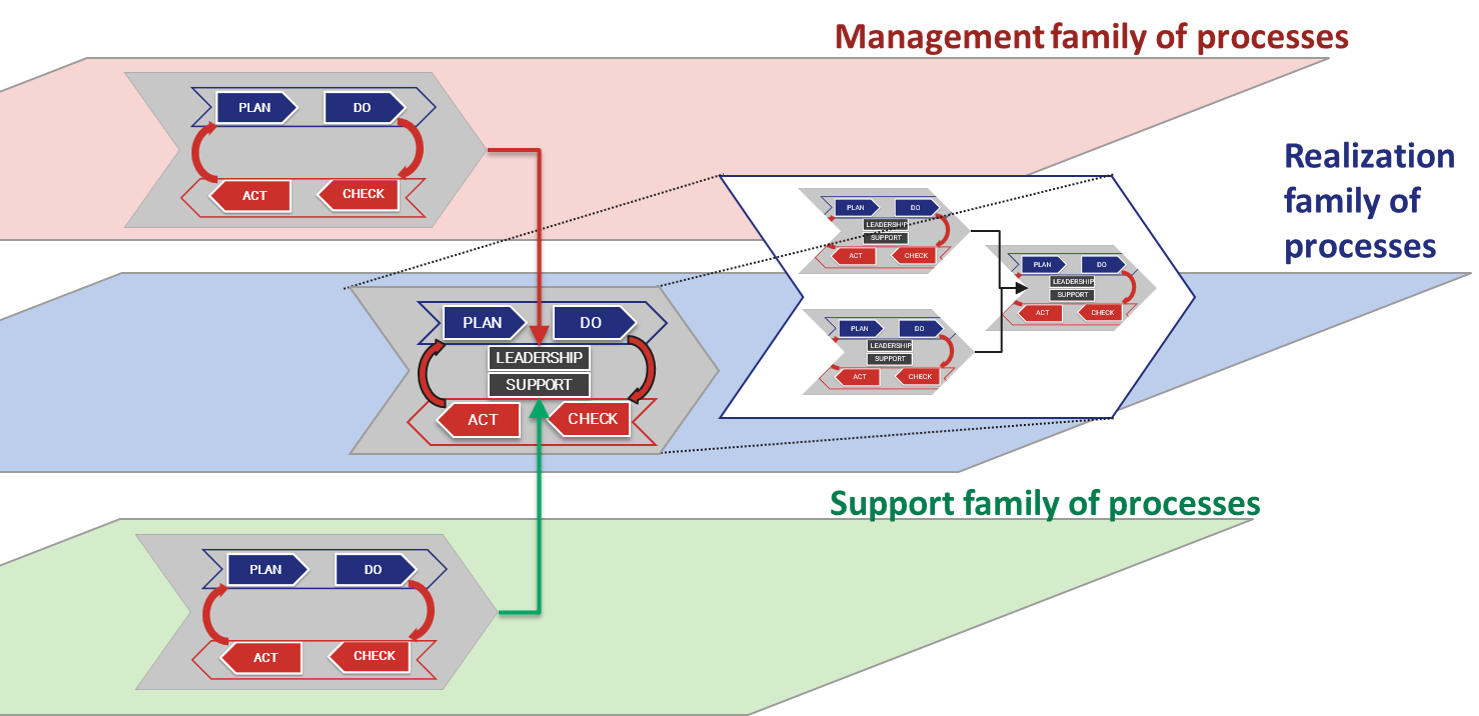}
    \caption{Generic process map of a management system integrating \pdca at each process and activity level}
    \label{ArtApiaCartoGenerique}
\end{figure}

This leads to the generic representation of an integrated management system given in \fig{ArtApiaCartoGenerique}, which we shall use later to integrate knowledge management processes. Note that the steering and support processes of a management system are full-fledged processes themselves. They also have their own steering and support sub-processes, which we've omitted here for simplicity.

\begin{figure}[ht]
    \centering
    \includegraphics[width=\textwidth/2]{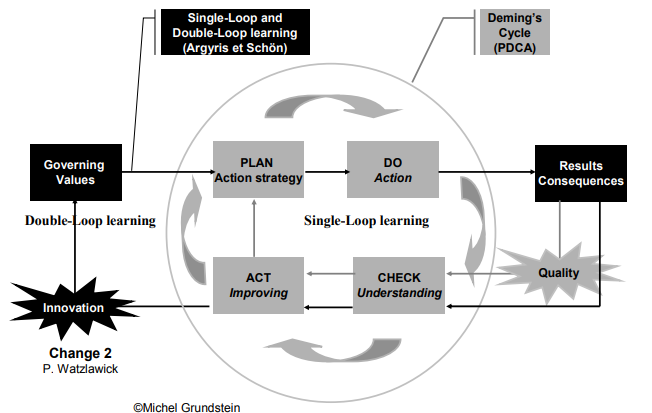}
    \caption{\sic{Deming’s cycle and Argyris \& Schön’s Organizational learning} as seen by Michel Grundstein in \cite{Grundstein2012} }
    \label{ArtApia2LoopMG}
\end{figure}

It is worth noting that Michel Grundstein \cite{Grundstein2012}, as early as 2012, established a connection between \lepdca (Plan-Do-Check-Act) as a driver for continuous organizational improvement and Knowledge Management (KM) as a driver for the learning organization. He observed that \lepdca simply and precisely corresponds to the concept of "single-loop learning" developed by Argyris and Schön \cite{Argyris2002}, while the ability to implement "double-loop learning", from the same authors, corresponds to, or at least integrates with, what is commonly called "innovation" within organizations. In this regard, Grundstein points out the convergence with the concept of "Change 2" (change in the internal composition law that governs the system) developed by Watzlawick et al. \cite{Watzlawick1975}. \fig{ArtApia2LoopMG}, extracted from Grundstein's article, illustrates this finding.

Our own observation is that, on one hand, when the \lepdca cycle is correctly practised, it effectively tends to ensure single-loop learning. It does this by surfacing corrective and preventive adjustments related to established knowledge and know-how that, through their use, have led to non-compliant or undesirable results. This, in turn, leads to updates in business knowledge repositories.

On the other hand, while we now systematically see innovation processes implemented to foster the emergence of incremental or disruptive innovations, we less frequently observe that these processes be nourished by an institutional capacity to systematically move beyond single-loop learning and enter into double-loop learning. This second loop aims to challenge the values and principles that led to the establishment of the knowledge and know-how that are being used. It is precisely this gap that, in our view, the knowledge management process can fill by deploying the \leseci model's mechanisms to complement the \lepdca cycles within the organization's processes.

\section{The SECI model (knowledge creation and application spiral }
In 1995, Ikujirō Nonaka et Hirotaka Takeuchi \cite{Nonaka95} proposed the \leseci model, a cycle for knowledge creation within an organization. This model consists of four mechanisms: Socialization « \seci{\bf S}{} », Externalization « \seci{\bf E}{} », Combination « \seci{\bf C}{} » and Internalization « \seci{\bf I}{} » of knowledge. The authors evolved their model in 2019 \cite{Nonaka19} into a spiral of knowledge creation AND application (\sic{practice}) by introducing temporal and societal (\sic{ontological}) dimensions. They also positioned Aristotelian {\it phronesis} (\sic{practical wisdom}), present in every individual, as the driving force behind this spiral.
 
This model is built on the idea that knowledge emerges, circulates and is transformed within organizations as employees apply it. This happens through a process of converting individual tacit knowledge into organizational explicit knowledge, which is then shared collectively and so on.

\begin{figure}[ht]
    \centering
    \includegraphics[width=\textwidth/2]{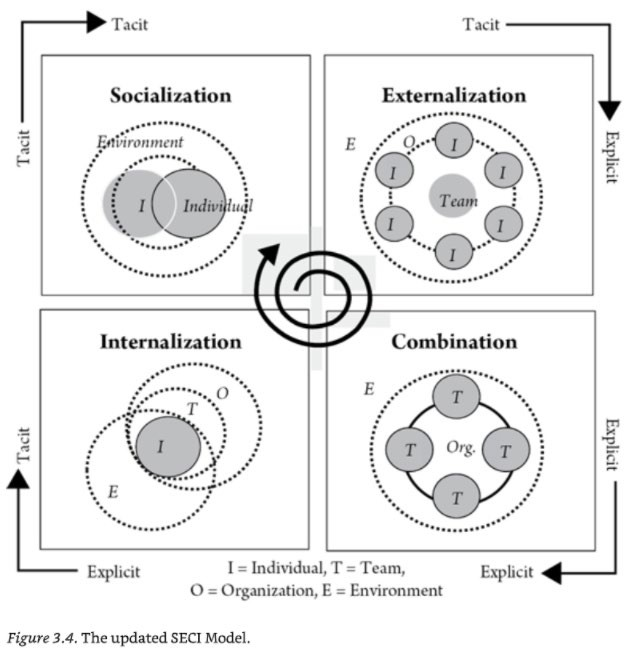}
    \caption{\sic{The updated \leseci ~Model}, from \cite{Nonaka19} }
    \label{ArtApiaSECI}
\end{figure}

 \fig{ArtApiaSECI}, from \cite{Nonaka19}, illustrates the four mechanisms of the \leseci  model. These correspond to specific combinations of the tacit or explicit forms knowledge takes and the vectors of this knowledge, ranging from individual to collective (individual, small group of individuals, whole organization).

 Socialization « \seci{\bf S}{} » and Internalization « \seci{\bf I}{} » primarily deal with the tacit knowledge individuals gain through personal experience or learning and which they enrich through exchanges with colleagues. 
  In contrast, Externalization  \seci{\bf E}{} » and Combination « \seci{\bf C}{} » involve the conversion of tacit knowledge into explicit knowledge which can thus be easily codified in the best adapted language (natural, mathematical, computer,...). Knowledge is made explicit, codified and validated collectively - by small groups of individuals, then globally by the organization - only to be eventually leveraged through its use by individuals in their operational situations after they have merged it with their own current knowledge through Internalization (formal and action learning).
While the literature contains many examples of \leseci model implementation, such as in \cite{Nishihara-2018CD} and \cite{Nishihara-2018PA}, we, for our part, based on the assessments we perform, have never truly encountered an organization in France that has institutionalized the application of this entire model. 
However, it is common to observe a partial deployment, particularly concerning the socialization and internalization mechanisms.

Examples include:
\begin{itemize}
\item Handover processes during an announced departure (often referred to as "biseau" in French, implying a smooth transition).
\item The existence of communities of practice dedicated to themes important to the organization.
\item The presence of expertise directories aimed at finding and soliciting the right expert when a problem arises.
\item Mentorship or coaching for new employees during their onboarding.
\end{itemize}

\begin{figure}[ht]
    \centering
    \includegraphics[width=\textwidth/2]{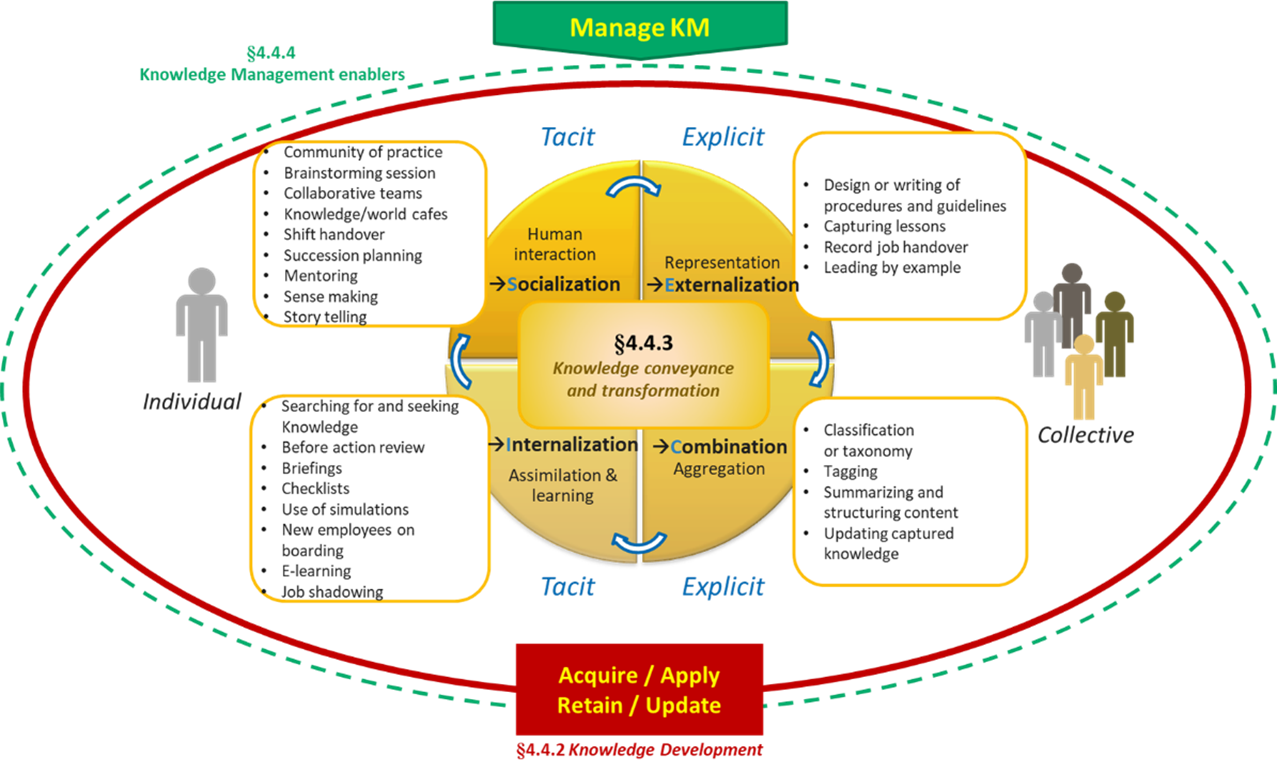}
    \caption{Ardans' view of \leseci transformation and conveyance mechanisms in §4.4.3 of \isoKM with examples of suited tools, methods and behaviours} 
    \label{Fig_ISO}
\end{figure}

\isoQ, in its paragraph 7.1.6, required organizations to demonstrate their control over "organizational knowledge," defined as "knowledge necessary for the operation of its processes and to achieve conformity of products and services." Many organizations, unsure how to meet this requirement, did face non-conformities during re-certification audits. The \leseci model, standing behind clause 4.4.3 of \isoKM offered a solution to meet \isoQ 7.1.6 requirement. It clarifies which knowledge needs to be managed: tacit, explicit, individual and collective knowledge. It also lists a number of tools, methods and behaviours favourable to implementing each \leseci mechanism. In \fig{Fig_ISO}, we reposition the "activities and behaviors" listed by \isoKM in relation to the \leseci  model.

Nevertheless, the translation of the \leseci mechanisms, whose presence and effects we ultimately wish to observe within an organization, into deployable, objective processes (as listed in clause 4.4.2 of the standard) remains unclear in \isoKM.

\section{Integrating Knowledge Management within an Integrated Management System through combining PDCA cycle and SECI model}

Through several examples of knowledge management system implementations in an ISO9001-compatible environment, we observed that the \lepdca cycle and the \leseci model could be naturally aligned and combined.

\subsection{The SECI model at work within the « Plan » (P) and « Do » (D) stages of the PDCA cycle}
These first two stages, which are inherently operational and value-creation oriented, require individuals to have the necessary knowledge to effectively prepare and execute their value-creating activities. It's therefore essential to empower these individuals by fostering the «~\seci{\bf I}{}~» (\seci{I}{nternalization}) and «~\seci{\bf S}{}~» (\seci{S}{ocialization}) aspects of the \seci model. These are reflected in the "assimilation and learning" and "human interaction" described in clauses 4.4.3 a and b of \isoKM. This implies establishing and maintaining technological and social environments that guarantee the accessibility, sharing and diffusion of relevant knowledge, whether explicit or tacit, thereby enabling individuals to effectively prepare and perform their work.

\begin{itemize}
\ittguilli{\seci{\bf I}{}nternalization}{(\fig{Fig_ArtApia_07}) Individuals learn from explicit organizational knowledge shared through documents (standards, operating procedures, rules, repositories, specifications, etc.), but also via online learning systems or with the help of recognized trainers. They convert this new explicit knowledge into tacit knowledge through their personal experience of applying this knowledge in a given professional context. This step allows the organization to verify whether its collective explicit knowledge capital is effectively available and applied by its employees when carrying out professional activities.}
\begin{figure}[ht]
    \centering
    \includegraphics[width=\textwidth/2]{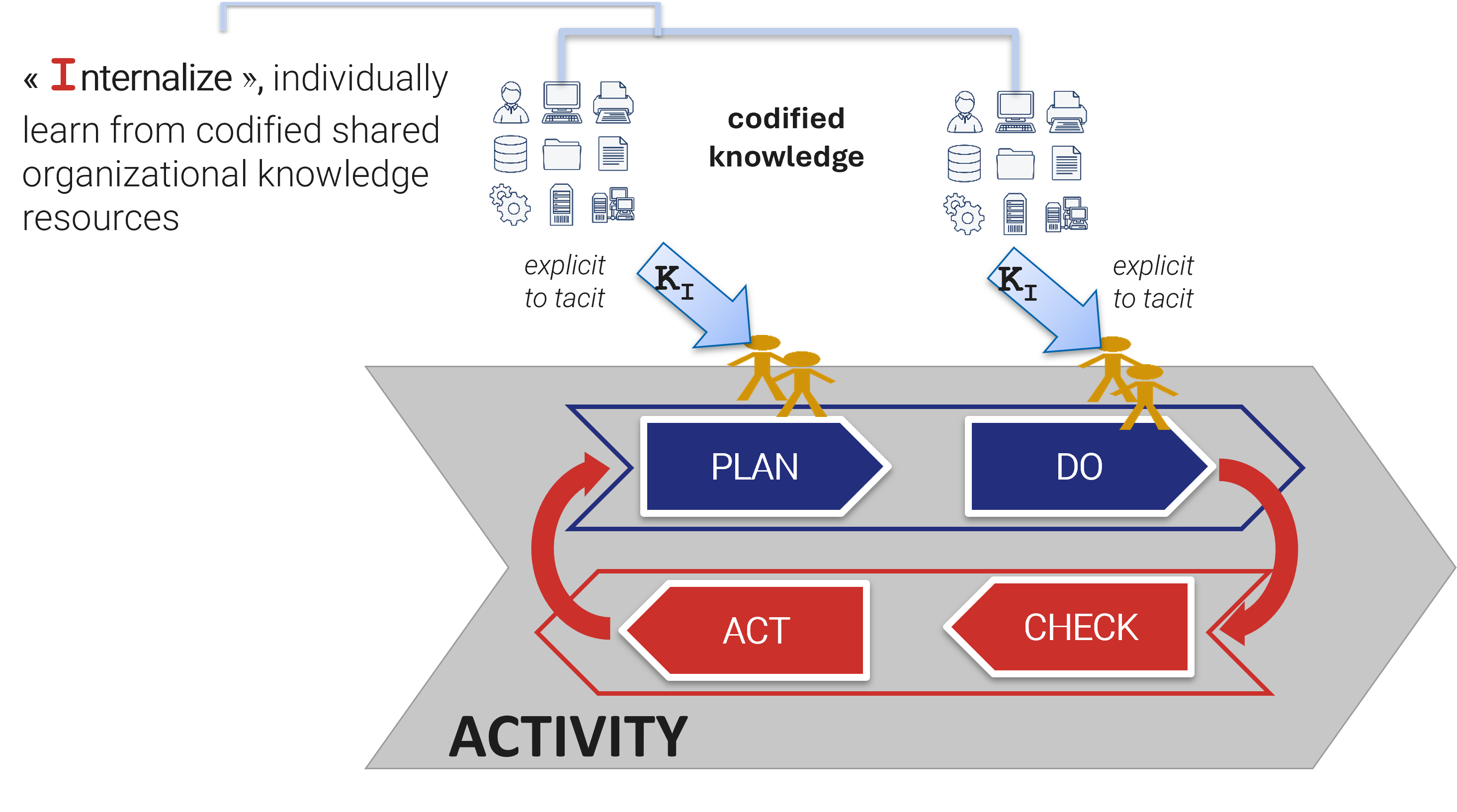}
    \caption{\seci{I}{nternalization} and \lepdca}
    \label{Fig_ArtApia_07}
\end{figure}

\ittguilli{\seci{\bf S}{}ocialization}{(\fig{Fig_ArtApia_08}) While carrying out their operational activities (Planning and Doing), individuals learn, compare and generate new insights and knowledge by sharing tacit knowledge through direct interactions, particularly within cross-functional teams (e.g., pairing and mentorship), communities of practice, or professional networks. This step also allows the organization to ensure its collective tacit knowledge assets are made available and utilized, even if tracking this effect is more complex.}
\begin{figure}[ht]
    \centering
    \includegraphics[width=\textwidth/2]{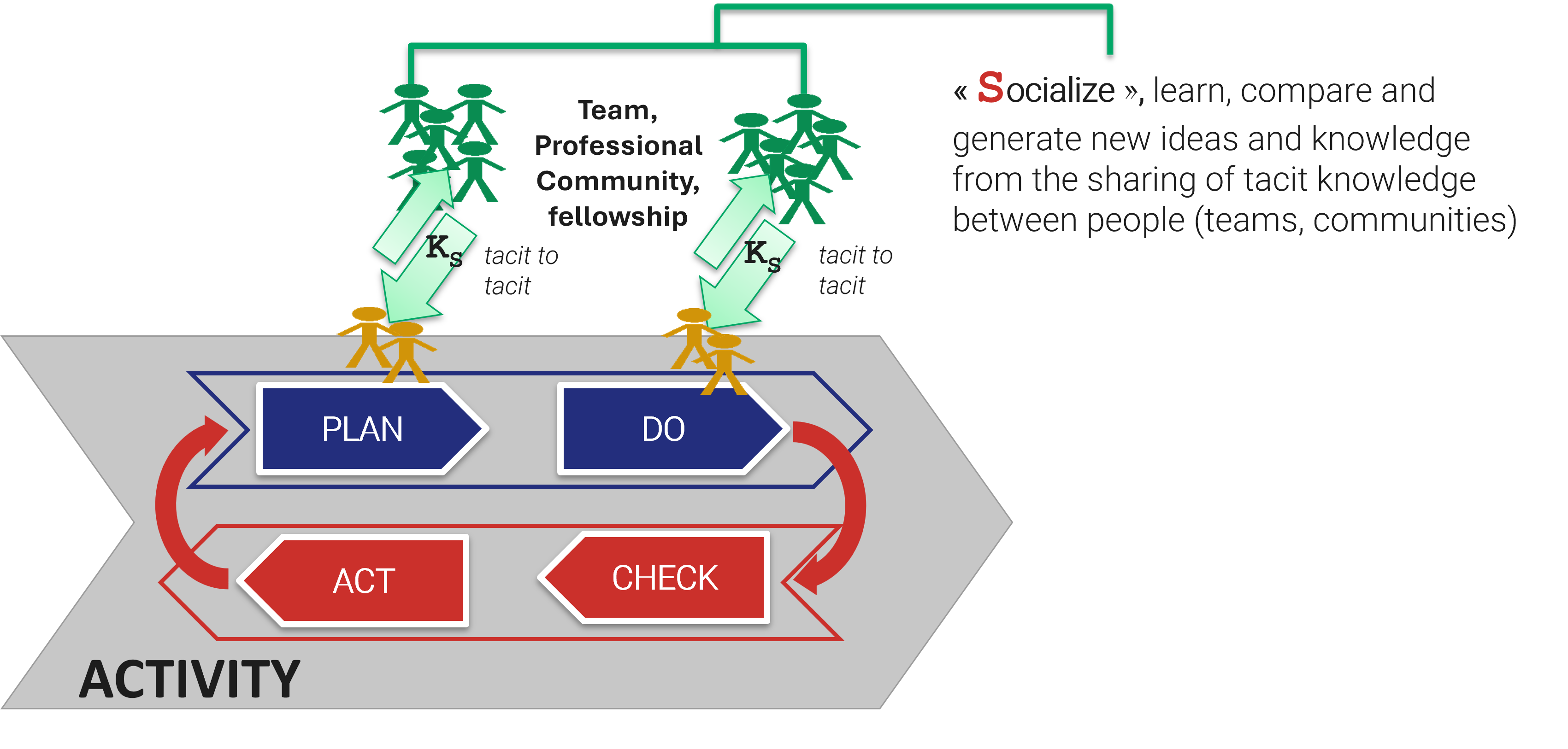}
    \caption{\seci{S}{ocialization} and \lepdca}
    \label{Fig_ArtApia_08}
\end{figure}
\end{itemize}

\subsection{The SECI Model in the "Check" (C) and "Act" (A) Stages of the PDCA Cycle}
Unlike the first two stages (Plan and Do) of \lepdca which are operational, stages C and A (Check and Act) are reflective steps aimed at observing, questioning - confirming or invalidating - and improving the way the Plan and Do stages were carried out, hence directly targeting and challenging the existing knowledge that was applied in these stages. They also involve identifying newly created knowledge in order to make this new knowledge explicit and thus formally shareable at the scale of the organization.

The «~\seci{\bf E}{}~» \seci{E}{xternalization} and «~\seci{\bf C}{}~» \seci{C}{ombination} mechanisms of the \leseci model are fairly equivalent to the Check «~\pdca{\bf C}{}~» and Act «~\pdca{\bf A}{}~» steps of the \lepdca cycle provided that these latter steps are performed in a "knowledge-oriented" manner.  Based on the observed results of a task, any challenged existing knowledge and any newly created knowledge is to be identified, made explicit, validated and combined with existing knowledge resources. This continuously enriches the organization's knowledge bases.

\begin{itemize}
\ittguilli{{\bf E}xternalization}{(\fig{Fig_ArtApia_09}) Knowledge used during planning and doing, whether derived from existing potential knowledge or created {\it de facto}, is revealed, supplemented and made explicit within its context. Indeed, with the "Check" (~\pdca{\bf C}{}~») stage, individuals collectively reflect on their own experiences and the observed results of applied tacit or explicit knowledge. This is done to understand and formalize new knowledge or improve existing knowledge (objective of the "Act" («~\pdca{\bf A}{}~») stage), such as for instance the building of a consensual understanding of the root causes of a failure or a success. This type of exercise is commonly practiced and called \sic{experience feedback} or \sic{lessons identified / lessons learned}.}
\begin{figure}[ht]
    \centering
    \includegraphics[width=\textwidth/2]{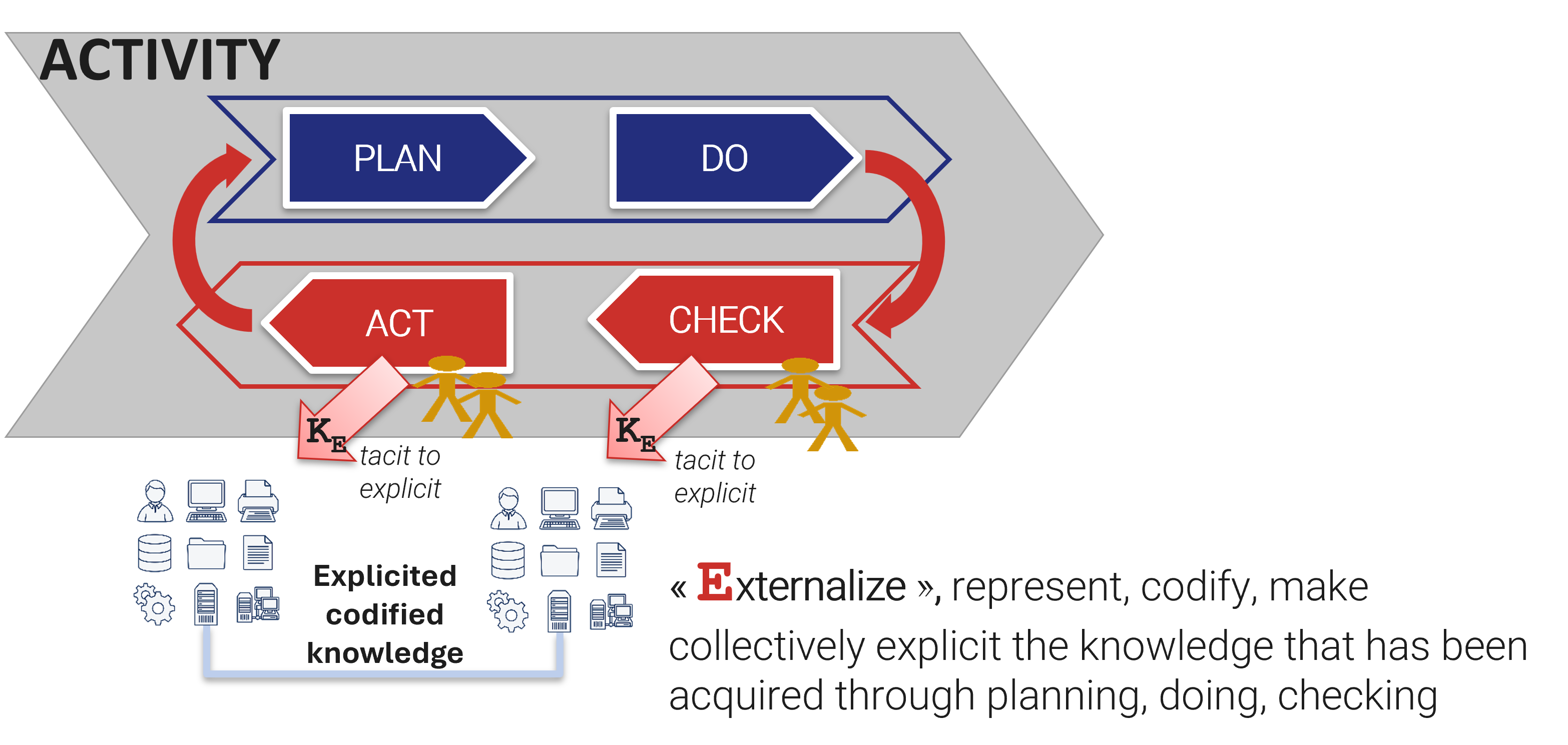}
    \caption{\seci{E}{xternalization} and \lepdca}
    \label{Fig_ArtApia_09}
\end{figure}
\ittguilli{{\bf C}ombination}{(\fig{Fig_ArtApia_10}) Knowledge is elevated to the rank of applicable, collectively organized and validated knowledge. The fractal principle of the model (i.e. the same principles are applied at the task, activity and process levels) ensures that, over time, knowledge transitions from being field-learned lessons or created knowledge, to knowledge that is validated and standardized  by the business lines and best practices recognized at the highest level (\sic{Corporate Knowledge of the Art}).}
\begin{figure}[ht]
    \centering
    \includegraphics[width=\textwidth/2]{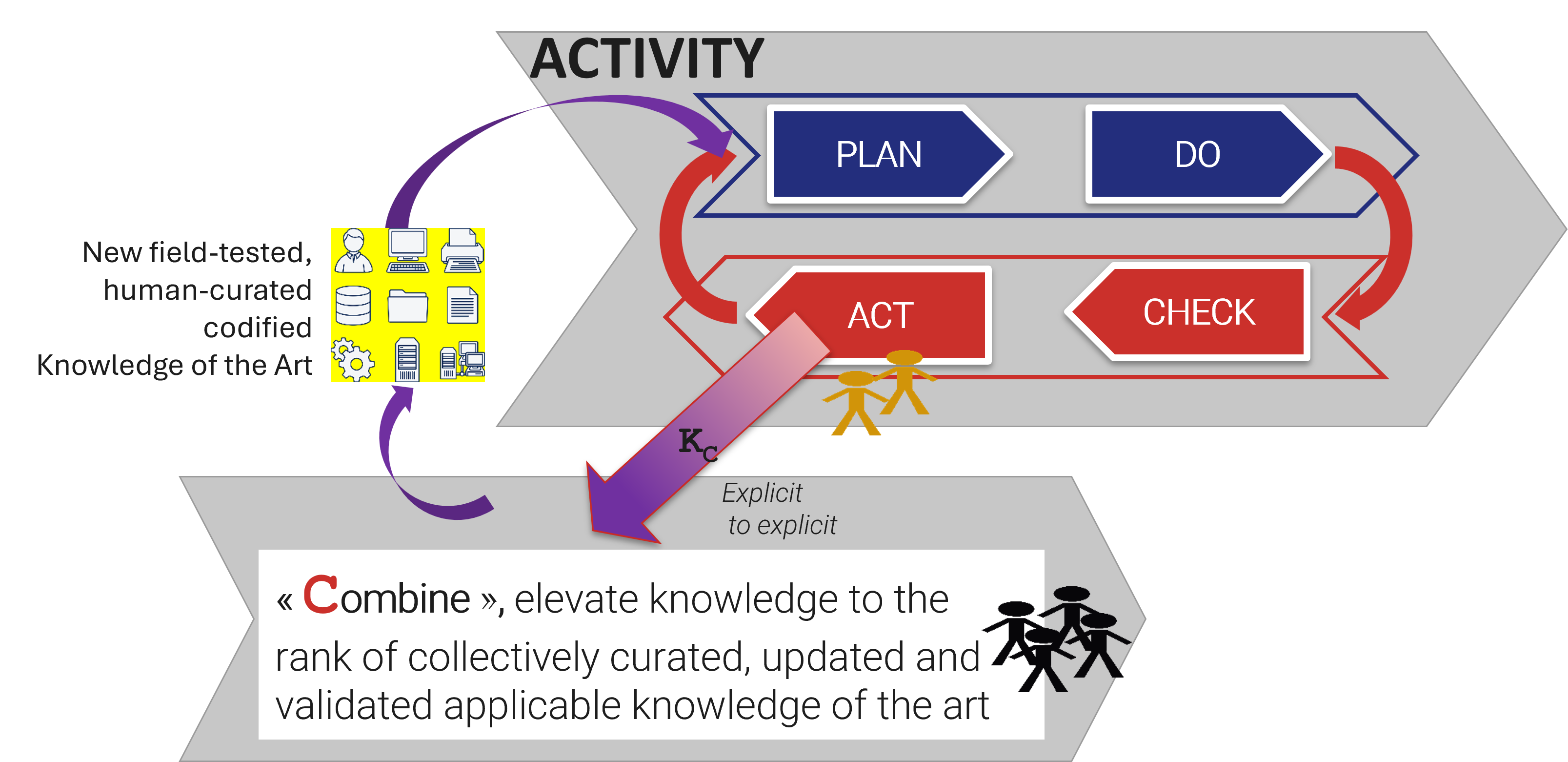}
    \caption{\seci{C}{ombination} and \lepdca}
    \label{Fig_ArtApia_10}
\end{figure}
\end{itemize}

Building on these four convergences between the \lepdca cycle and the \leseci model, we can translate this concretely into a holistic vision of the knowledge management system integrated within an IMS (Integrated Management System), provided that two complementary and essential processes are taken into account: leadership and support for knowledge management activities, as depicted in the next section.

\section{Holistic Vision of the Knowledge Management System processes}

To align with the systemic view of the process approach, an organization's integration of a Knowledge Management System into its Integrated Management System (IMS) requires, beyond what we have described above at the realization level, the taking into account of two complementary sub-processes: knowledge management leadership and support.

\subsection{Knowledge Management Leadership process}
This process, belonging to the "Leadership" process family, focuses on developing the KM action plan and overseeing its implementation. This process includes the following activities:
 \begin{itemize}
\ittguilli{Establishing a KM Policy}{Defining the organization's approach to knowledge creation, sharing and use, aligning it with the organization's objectives and strategy.}
\ittguilli{Developing and Implementing a Knowledge Governance Framework}{Designating the organizational structure, roles and responsibilities to ensure consistent application of knowledge management principles at all levels and functions.}
\ittguilli{Integrating \leseci Model Principles into the \lepdca Cycle}{Recognizing that the four steps – socialization, externalization, combination and internalization – are essential throughout the entire \lepdca cycle, without being limited to specific stages.}
\ittguilli{Defining a KM Action Plan}{This plan aims at translating the policy, particularly by developing a map of key and vulnerable knowledge to identify and prioritize actions to be taken.}
\ittguilli{Defining the cultural and technical transformation actions}{these are the actions necessary for the adoption of knowledge management practices.} 
\end{itemize}

\subsection{Knowledge Management Support Process}
This process, belonging to the "Support" process family, specifically involves:
\begin{itemize}
    \ittguilli{Providing KM methods and resources (human and technical) to activate the mechanisms for knowledge internalization and socialization)}{for instance, mentorship programs within communities of practice, smart search in business knowledge repositories, etc... This sub-process facilitates the creation, sharing and learning of knowledge by individuals, the integration of existing knowledge into decision-making processes and the monitoring of its application at all \lepdca stages, especially "PLAN \& DO," across all targeted business processes and activities.}
    \ittguilli{Providing KM methods and resources (human and technical) to activate knowledge externalization and combination mechanisms }{for instance, project lessons learned workshops, expert knowledge modelling within communities of practice, populating business knowledge repositories, etc... This facilitates the identification, articulation and combination, challenging, validation and structured storing of knowledge within the "CHECK \& ACT" stages of all targeted business processes and activities.}
\end{itemize}

\begin{figure}[ht]
    \centering
    \includegraphics[width=\textwidth/2]{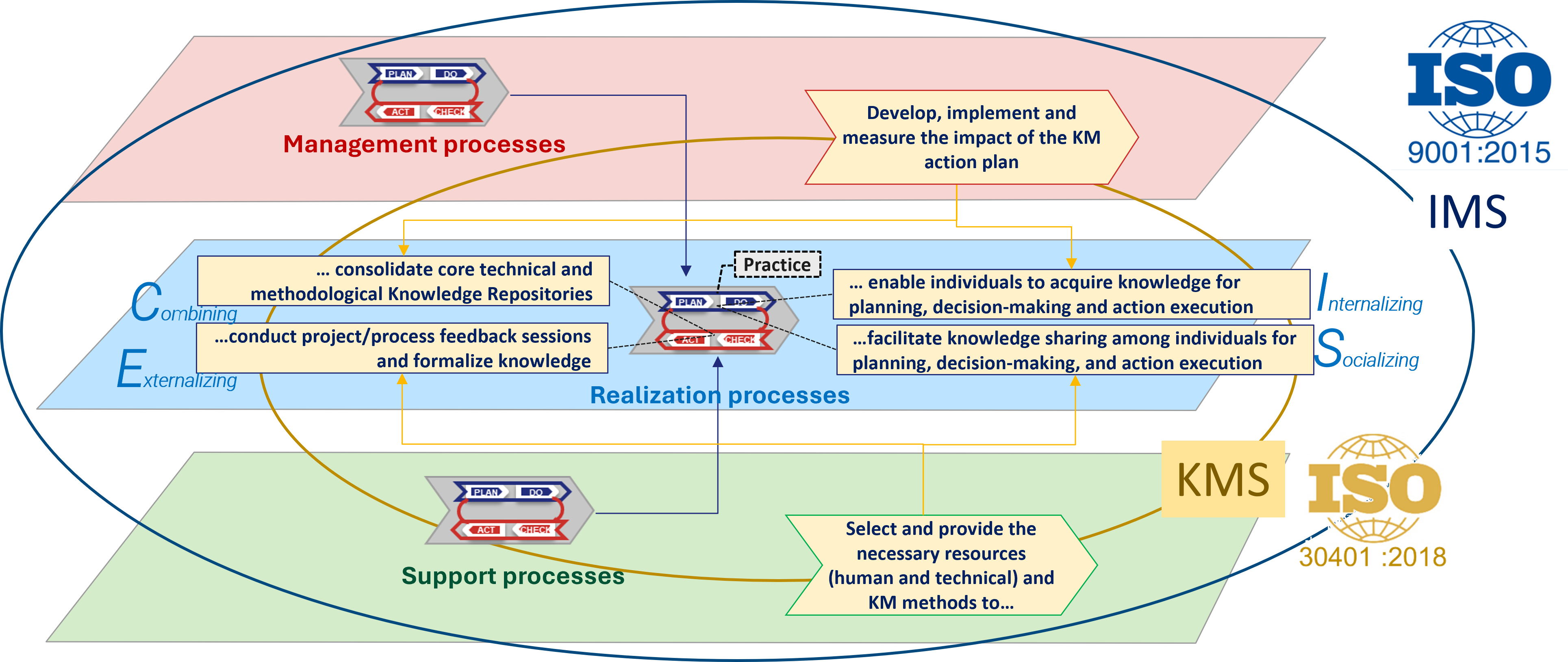}
    \caption{Integrating a KMS into an IMS}
    \label{Fig_ArtApia_11}
\end{figure}

\fig{Fig_ArtApia_11} summarizes the positioning of these KM management and support processes and their interactions with the deployment of the \lepdca cycle within operational processes. This leads to the integration of activities and behaviours derived from the \leseci model at each stage of the \lepdca cycle:

\begin{itemize}
    \ittguillj{Individual appropriation}{of knowledge made available by the operational teams;}
    \ittguillj{Knowledge sharing} {among operational personnel : peers from the same discipline or, conversely, practitioners from complementary disciplines;}
    \ittguillj{Explicit articulation}{of collectively acquired knowledge through exercises like "lessons learned" (REX) or expertise gathering, which is the responsibility of the operational staff who experienced the origination of such new or updated knowledge;}
    \ittguillj{Consolidation}{of process action frameworks and business repositories by integrating this new knowledge (impact analysis on existing explicit knowledge, updating classification structures, etc.), which is the responsibility of process owners and business managers.}
\end{itemize}

\section{Conclusion}
Following the introduction of the requirement to manage organizational knowledge in \isoQ, many companies, certified or not, asked themselves \sic{How do we go about it?}. Too often, we have seen a mere "KM process" appear in the process maps of Management Systems, categorized under Support processes, supposedly providing help with knowledge capitalization and dissemination.
Given this observation, we have shown that KM is indeed a system comprising a leadership process, a support process and the knowledge management realization process itself, the latter consisting of activities intimately intertwined with the organization's operational processes.
While the leadership and support processes for knowledge management — leading to the establishment of governance, methods and environments conducive to organizational learning — fall under KM professionals, the knowledge management process itself falls under operational staff who create, update and make use of this knowledge. 
Our conviction, based on the examples we have handled in recent years, is that this process can be implemented quite naturally, even "painlessly" in terms of effort required, by applying the \leseci model for knowledge creation and application, especially when the \lepdca cycle is a well effective practice for all employees within the organization

\bibliographystyle{plain}
\bibliography{Biblio_Ardans_SKM_SMI}

\end{document}